\documentclass[conference]{IEEEtran}
\pdfoutput=1
\IEEEoverridecommandlockouts
\usepackage{cite}
\usepackage{amsmath,amssymb,amsfonts}
\usepackage{algorithmic}
\usepackage{graphicx}
\usepackage{textcomp}
\usepackage{xcolor}
\usepackage{subcaption}

\def\BibTeX{{\rm B\kern-.05em{\sc i\kern-.025em b}\kern-.08em
    T\kern-.1667em\lower.7ex\hbox{E}\kern-.125emX}}

\begin{document}

\title{How to Make Your Multi-Image Posts Popular? An Approach to Enhanced Grid for Nine Images on Social Media}

\author{
    \IEEEauthorblockN{Qi Xi\IEEEauthorrefmark{1,*}, Shulin Li\IEEEauthorrefmark{1,*}, Zhiqi Gao\IEEEauthorrefmark{2}, Zibo Zhang\IEEEauthorrefmark{1}, Shunye Tang\IEEEauthorrefmark{1}, \\Jingchao Zhang\IEEEauthorrefmark{1}, Liangxu Wang\IEEEauthorrefmark{1}, Yiru Niu\IEEEauthorrefmark{3}, Yan Zhang\IEEEauthorrefmark{1}, Binhui Wang\IEEEauthorrefmark{1,4}\textsuperscript{*}}
    \IEEEauthorblockA{
        \IEEEauthorrefmark{1}College of Software, Nankai University, Tianjin, China \\
        \{2314344@mail.nankai.edu.cn, lishulin0119@mail.nankai.edu.cn, zhangzibo@mail.nankai.edu.cn, \\
        2113298@mail.nankai.edu.cn, 2111636@mail.nankai.edu.cn, 2113198@mail.nankai.edu.cn,\\
        2012151@mail.nankai.edu.cn,
        wangbh@nankai.edu.cn\}
    }
    \IEEEauthorblockA{\IEEEauthorrefmark{2}Business School, Nankai University, Tianjin, China \\
    gaozhiqi@mail.nankai.edu.cn}
    \IEEEauthorblockA{\IEEEauthorrefmark{3}School of Sociology, Nankai University, Tianjin, China \\
    2110231@mail.nankai.edu.cn}
    \IEEEauthorblockA{\IEEEauthorrefmark{4}Tianjin Key Laboratory of Software Experience and Human Computer Interaction, Tianjin, China \\
    wangbh@nankai.edu.cn}
}

\IEEEoverridecommandlockouts
\IEEEpubid{\makebox[\columnwidth]{978-1-7281-8156-0/20/\$31.00 ©2020 IEEE \hfill} \hspace{\columnsep}\makebox[\columnwidth]{ }}
\IEEEpubidadjcol

\maketitle

\begin{abstract}
The nine-grid layout is commonly used for multi-image posts, arranging nine images in a tic-tac-toe board. This layout effectively presents content within limited space.Moreover, due to the numerous possible arrangements within the nine-image grid, the optimal arrangement that yields the highest level of attractiveness remains unknown. Our study investigates how the arrangement of images within a nine-grid layout affects the overall popularity of the image set, aiming to explore alignment schemes more aligned with user preferences. Based on survey results regarding user preferences in image arrangement, we have identified two ordering sequences that are widely recognized: sequential order and center prioritization, considering both image visual content and aesthetic quality as alignment metrics, resulting in four layout schemes. Finally, we recruited participants to annotate various layout schemes of the same set of images. Our experiential-centered evaluation indicates that layout schemes based on aesthetic quality outperformed others. This research yields empirical evidence supporting the optimization of the nine-grid layout for multi-image posts, thereby furnishing content creators with valuable insights to enhance both attractiveness and user experience.
\end{abstract}

\begin{IEEEkeywords}
Multi-image posts, Image Popularity,Nine-grid Layout,Online Sharing,Visual Attention
\end{IEEEkeywords}

\section{Introduction}
The widespread popularity of multi-image posts on social media platforms has reshaped the way content is consumed and shared online. Among various formats, the nine-grid layout stands out as a prominent and versatile option for presenting multiple images cohesively within a single post. This layout arranges nine images in a three-by-three grid, offering a visually engaging narrative that captures users' attention and encourages interaction. The distinct characteristics and advantages of the nine-grid layout have led to its extensive use across personal and commercial contexts, serving as a powerful tool for storytelling, branding, and marketing.
\begin{figure}
    \centering
    \includegraphics[width=0.37\textwidth]{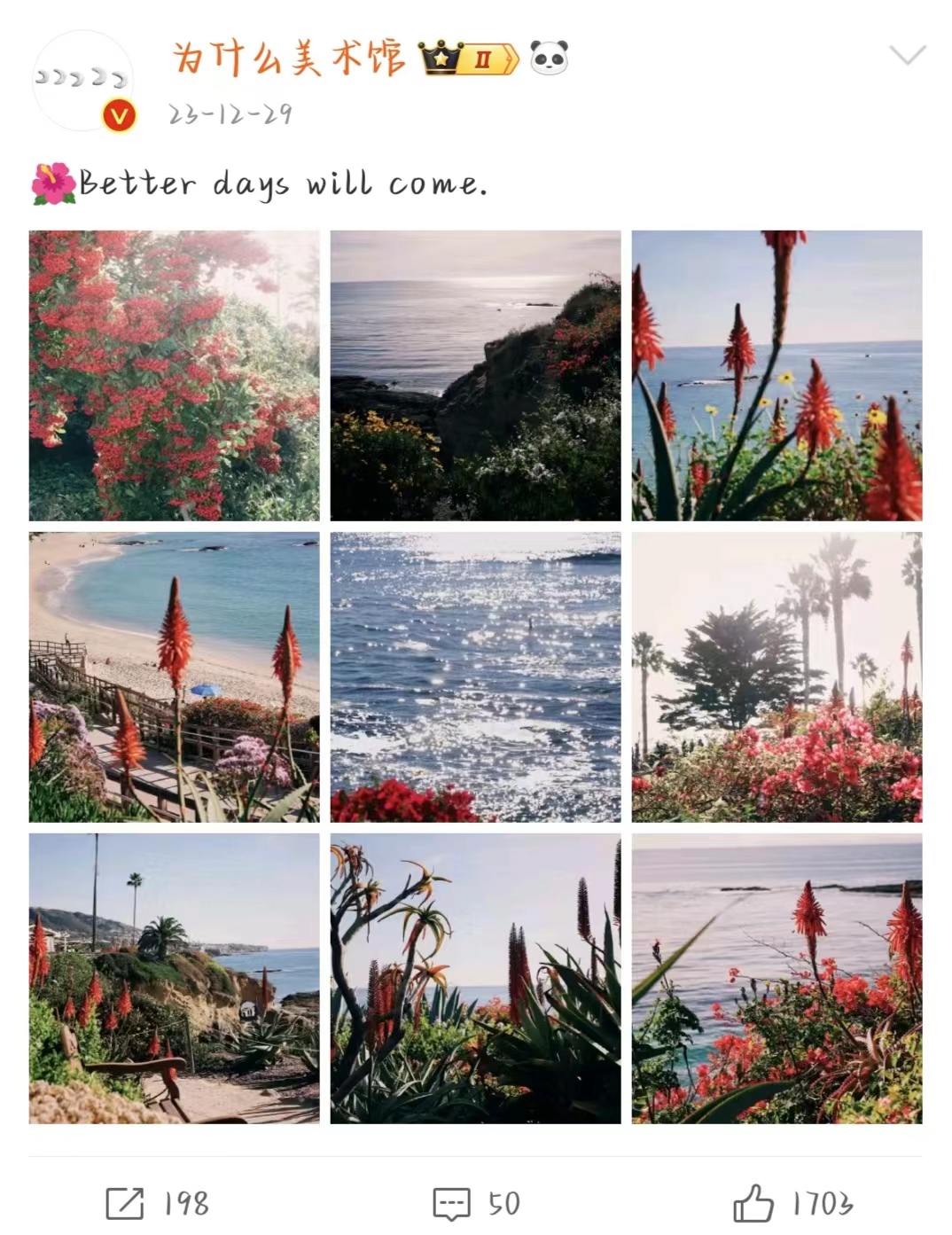}
    \caption{An example of a multi-image post with nine-grid layout}
    \label{fig:enter-label}
\end{figure}

However, despite its prevalence and effectiveness, the optimal arrangement of images within the nine-grid layout remains a challenge. The arrangement of images plays a crucial role in determining the overall appeal and effectiveness of the post. Thus, the question arises: How should images be arranged within the nine-grid layout to maximize engagement and aesthetic appeal?

Existing research in this area has primarily focused on image evaluation and popularity prediction in single-image contexts, overlooking the complexities of multi-image layouts such as the nine-grid. Consequently, there is a gap in understanding how different arrangements of images impact the overall attractiveness and user engagement of multi-image posts.

Our study addresses this gap by proposing several arrangement schemes for the nine-grid layout and exploring automated arrangement techniques to streamline the process. Our research aims to contribute to the field by investigating more attractive and effective arrangement strategies tailored specifically for multi-image posts. We introduce a novel approach called Optimal Grid for Multiple Images (OGMI), which offers customizable arrangement options based on image content, aesthetics, and user preferences.

To evaluate the effectiveness of our proposed solutions, we conducted experiments using a user experience evaluation tool integrated into a web-based platform. Through these experiments, we aim to uncover insights into users' preferences and perceptions regarding different arrangement schemes, both in personal and commercial contexts.

In summary, this study presents a comprehensive exploration of the nine-grid layout's arrangement problem in the context of multi-image posts. By introducing innovative solutions and conducting empirical evaluations, we aim to advance understanding and offer practical guidance for creators and marketers seeking to optimize the visual presentation of multi-image content on social media platforms.

\section{Background and Related work}

\subsection{Background}

The nine-grid layout, consisting of a 3x3 grid of squares, is a common pattern widely used in various domains. In today's social media, people are enthusiastic about showcasing images in a nine-grid format to express their appreciation for things and their love for life. Additionally, the nine-grid layout continues to find extensive applications in other fields such as user interfaces of mobile devices, game design, and information display. The arrangement of the nine-grid has a significant impact on user experience and usability. An effective arrangement can enhance user efficiency, visibility of information, and overall aesthetic appeal, thereby making the nine-grid more popular. 

The nine-grid layout visually presents integrity, which is a special aspect of positioning images. Although there have been many studies on the impact of the order of images on user attention. However, there is still a lack of research on whether the order of images affects the user's overall liking in the nine-grid situation. The following study will be based on two existing theories of the position significance of images to study the most popular image layout method among users in the nine-grid situation.

\subsection{Image Popularity on social media}

In examining the influence of different arrangement methods on the nine-grid layout, we have reviewed several relevant studies. Firstly, Keyan Ding et al\cite{ding2019intrinsic}. aimed to develop a computational model for accurately predicting the viral potential of social images. Their study involved constructing a large-scale image database and employing a deep neural network-based computational model to analyze the individual contributions of image content to its popularity. However, their research primarily focused on scoring and predicting the popularity of single images, which differs from the focus of our study.Additionally, Hossein Taleb\cite{talebi2018nima} proposed a method based on deep object recognition networks to reliably assess image quality with high correlation to human perception. This approach was employed for predicting image quality assessment and shares some similarities with our study, although the emphasis is different.

Furthermore, Ma Xiaoyue et al\cite{ma2021image}. investigated the impact of image positioning and layout on user engagement behavior using multi-image tweet data from the Sina Weibo platform. They utilized an XGBoost model to predict the user engagement potential of images and conducted correlation and regression analyses, revealing that image positioning and layout moderately influence user interaction. This study explored the effects of different arrangement methods for multiple images, which overlaps to some extent with our research. Zhonghua He et al\cite{ma2023suggest}. explored the relationship between the sequential position of images in nine-image grid and the Big Five personality traits. Their findings demonstrated that users can utilize prominent positions in the nine-image grid to highlight specific content, resulting in more captivating Weibo narratives. This study enriched the field of social media image research and revealed motivations for using multiple images.

In contrast, our paper specifically focuses on examining the impact of different arrangement methods on the nine-grid layout. Our novelty lies in exploring the influences of various arrangement methods on the visual appeal and narrative expression of the nine-grid images. By comparing different arrangement methods, we aim to provide specific guidance and insights on the optimal arrangement strategies for the nine-grid layout. Our research fills a gap in the existing knowledge and offers practical implications for designers and practitioners seeking to optimize the arrangement of images in the nine-grid layout.

\subsection{Position significance of images}

There are two views on the position significance of images. The first one is the serial-position effect. It is the tendency of a person to recall the first and last items in a series best, and the middle items worst\cite{colman2015dictionary}. It was first documented by Ebbinghaus\cite{ebbinghaus1885gedachtnis} in his seminal work on memory in the late 19th century, which is a fundamental principle of cognitive psychology. Numerous studies have affirmed its robustness across various tasks and stimuli types. While most research has focused on verbal materials, recent studies have extended this phenomenon to visual stimuli, including images.

The serial-position effect manifests as two distinct phenomena: the primacy effect and the recency effect. The primacy effect refers to the enhanced recall of items presented at the beginning of a sequence. This phenomenon is thought to occur because early items receive more rehearsal and encoding time, allowing them to be transferred into long-term memory more effectively. Meanwhile, the recency effect pertains to the superior recall of items presented at the end of a sequence. This effect is believed to result from items being held in short-term or working memory, which is still active during recall tasks\cite{glanzer1966two}. In the context of image sequences (or nine-grid layout), the serial-position effect can be used to enhance users’ attention by strategically positioning impactful pictures at the beginning and the end. This balanced approach ensures that important information is not only introduced early but also reinforced at the end, maximizing the likelihood of retention and recall. 

The second definition of significant position of nine-grid layout adheres to visual attention distribution\cite{ma2023suggest}. Some researchers suggest that users will first pay attention to the center of the screen, which makes the position in the middle of the image sequence important. The image that is most compelling or fits the purpose most can be put in the center to gain more visual attention from users. However, further studies on which of the two definitions has the highest level of attractiveness in the situation of nine-grid layout are still needed. 

\section{Formative Study}

\subsection{Questionnaire Survey}

We conducted a survey to investigate the perception of social media platform users towards the nine-grid layout. Out of the 36 valid responses received, all participants indicated that they had encountered nine-grid image posts on social media platforms, with 75\% of them stating that they frequently browse such posts.

Subsequently, based on whether the participants had experience in creating nine-grid posts, the respondents were divided into two categories. The group with editing experience accounted for a high proportion of 77.8\%, among which 96\% of the respondents believed they possessed a certain level of arranging ability, while 93\% stated that arranging would consume a significant amount of their time.We further investigated the preferences and experiences of this group in arranging the nine-grid layout,and the statistical results are shown in the Figure 2.The survey results revealed that for the most satisfying image, people tended to prefer placing it in the P5 position (96.43\%) and the P1 position (32.14\%), with positions in the four corners (P1, P3, P7, P9) being more popular than those on the edges. As for the least satisfying image, it was widely accepted to place it in the P7 position (39.2\%), the P8 position (28.57\%), and the P9 position (50\%).For respondents who only had browsing experience, we attempted to understand their level of attention towards different positions in the nine-grid layout. According to statistics, the P5 position and the P3 position garnered significantly more attention than others, with probabilities of 100\% and 62.5\%.
\begin{figure}
    \centering
    \includegraphics[scale=0.5]{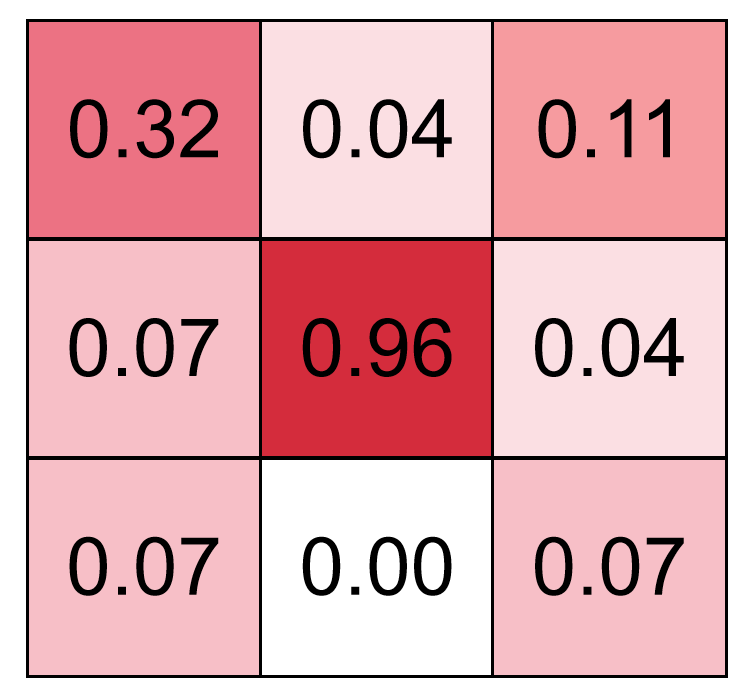}
    \caption{Vote on the position for the best image}
\end{figure}

\begin{figure}
    \centering
    \includegraphics[width=0.45\textwidth]{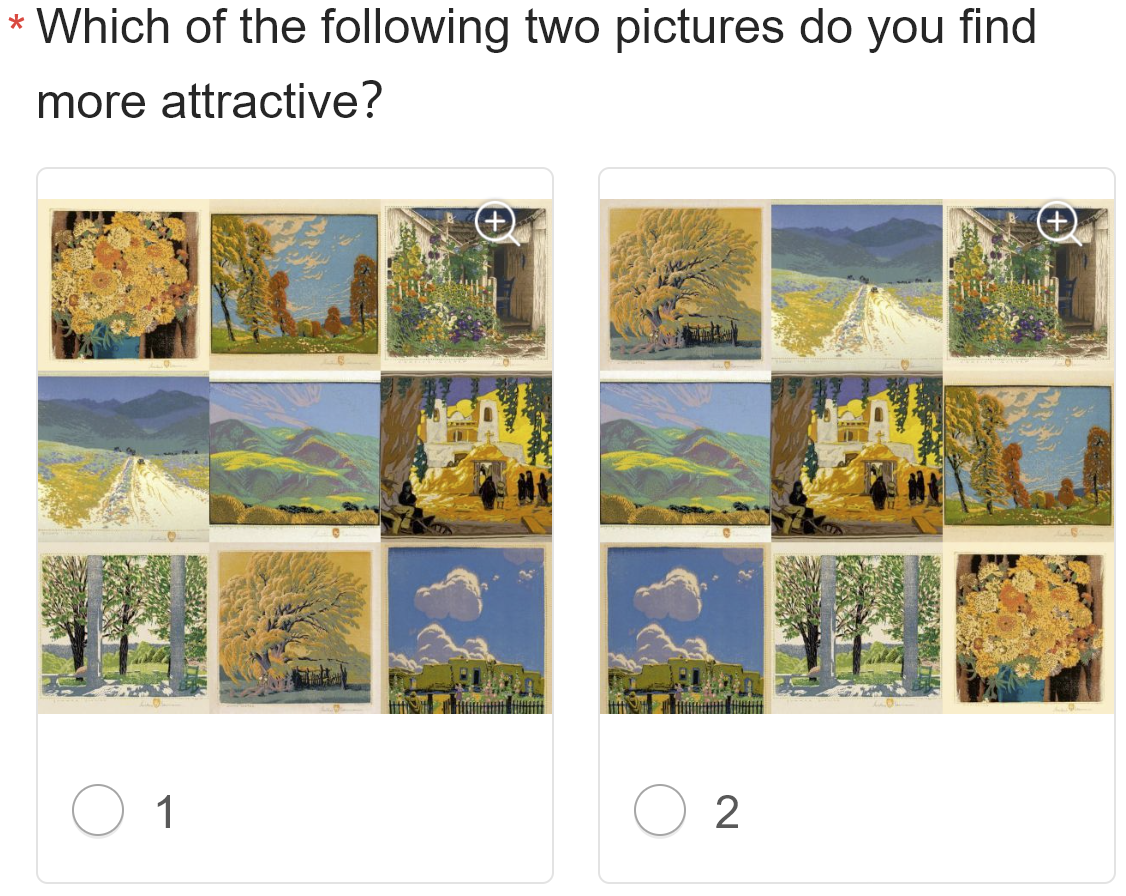}
    \caption{A set of image pairs from the questionnaire}
\end{figure}
Lastly, we constructed pairs of images consisting of a rearranged nine-grid thumbnail and its original arrangement. Participants were asked to choose the image with better presentation without knowing the labels of the unfamiliar images(Figure 2).This can help us to understand whether the arrangement of images is effective in improving the overall visual effect of the nine-grid images. The average number of votes for the ranked images was 23.4, significantly higher than the average votes received by the original images, which was 12.6.

\subsection{Semi-structured Interview}

To understand the factors that contribute to the popularity of nine-grid layout formats and explore the practices, preferences, and perceptions of individuals’ actions when utilizing the nine-grid layout, we conducted semi-structured interviews with five participants. By examining their current practices, image selection criteria, perceptions of prominence within the grid, and underlying motivations for employing this format, this study aims to lay the groundwork for devising effective strategies for enhancing the popularity and impact of multi-image content on social media platforms. 

We invited five individuals with prior experience on posting multi-image combinations on social media like Sina Weibo, WeChat Moments, Instagram, etc. to participate in semi-structured interviews. Each interview session began with participants presenting an example of their nine-grid layout, followed by a series of inquiries regarding their motivations for posting photos on social media, their typical posting routines, criteria for photo selection and evaluation, and their perspectives on the importance of specific positions within the nine-grid format. Finally, there was a free discussion part where they were encouraged to talk about their relevant experience and tricks they have used previously relating to this topic. Convenience sampling was adopted for participant selection, acknowledging the pragmatic nature of this sampling method within the scope of the study. No incentives or rewards were offered for their participation. Interviews were conducted individually via phone calls to accommodate participants' schedules and ensure the confidentiality of responses. 

\subsection{Findings}

\textbf{\emph{Purpose for posting multi-image posts on social media.}}
The primary purpose identified for posting multi-image posts on social media is documentation and sharing, as articulated by all participants, who emphasized terms such as “documenting life moments” and “sharing daily experiences.” Additionally, one participant highlighted the significance of making interpersonal connections through visually appealing images, viewing it as a means to engage with diverse audiences. The other saw this action as a way of expressing thoughts and emotions. One participant specifically underscored platform differences, noted that “posting original pictures on Weibo maintains their image quality, thus the purpose of this platform is primarily for archiving. On the other hand, posting on WeChat Moments is more about sharing daily mood and communicating with friends.” 
 
Moreover, participants expressed satisfaction in encountering multi-image posts created by others online. One person expressed their enjoyment, stating, “I enjoy watching imagery posts, especially when those pictures are arranged in an artful way. I followed a few bloggers who post photographs and movie screenshots every day as their posts entertain and inspire me aesthetically. I can really feel the content creator’s passion for making something beautiful if they pay extra attention on image arrangements.” 

\noindent\textbf{\emph{Current practice when posting images online.}}
Participants described their current practices when posting images online, outlining several key steps. Initially, they feel an impulse or need to create a post, often centered around a specific theme. Subsequently, they select images that align with this theme. Two participants said they would edit the selected photos to achieve a better artistic effect. Four out of five mentioned that they would arrange the images into a particular order before posting them. Additionally, two stressed the significance of crafting captions, as they would pay extra time to think of some texts matching the overall theme of the images. After all these processes, they post them onto the platform. 
 
When questioned further about their decision-making process for image arrangement, participants provided insights into their strategies. Two participants typically follow a chronological order, while another mentioned deviating from this order to create a compelling narrative behind the photos. One participant differentiated between two types of posts, detailing, “for travel logs, I organize photos in a chronological framework. While for pure aesthetic photos with person, they are arranged based on clothing colors or body postures.”

\noindent\textbf{\emph{Criteria for image selection.}}
Participants have outlined various criteria for selecting images, emphasizing aesthetic appeal, thematic relevance, and technical quality. Participants all believed that the selected images should fit with the overall theme and convey their intended messages. Three out of five participants cited aesthetic and quality factors as key considerations when assessing images, as they used phrases like “clarity”, “visually pleasing”, “harmonious color”, and “balanced composition”. One participant gave their priorities when selecting: appealing subject > captivating atmosphere > beautiful scenery. They added that if everything else in the picture was perfect except for their face, they would still post it while using photo-editing applications to attach a sticker over their face. In addition, the other participant mentioned deciding together with friends who also appear in the photo. If all were agreed that one image is good, then it shall be granted with a position in the post. 

\noindent\textbf{\emph{Most prominent position in a nine-grid layout format.}}
All the participants noted the central position (P5) among the most prominent ones. They described it as “visual center”, “core focus”, “set the tone for the whole post”. Three out of five participants also listed the top-left corner as important. This is because it is the first position seen by viewers following the reading order. They felt like if the image on this position is attractive, there would be a larger chance for the audience to scroll to the next picture, instead of leaving this stream. One person emphasized the importance of all four corners, as they would perceive them as a cohesive whole at first glance. P4 was mentioned once, and the participant added “it is my personal taste”. While the other one elaborated under which circumstances the bottom-right corner is significant for them: “if I have something to show off but don’t want to be too overt, I will place that photo to P9, where it may be noticed by only a minority of very patient viewers.”

\begin{figure}[h]
  \centering
  \includegraphics[scale=0.6]{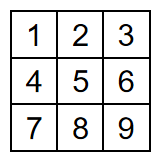}
  \caption{Illustration of position 1-9 in a nine-grid format.}
\end{figure}

\section{Methods}

\subsection{Intrinsic Image Popularity Assessment Model}

\subsubsection{Introduction to the $I^2PA$ Model}

Research in Automatic Image Popularity Assessment (IPA) aims to develop computational models capable of accurately predicting the viral potential of social media images. Existing popular methodologies, such as those proposed by Khosla\cite{khosla2014makes} and Hessel\cite{hessel2017cats}, typically integrate a blend of visual and non-visual factors, including user metrics and upload timing, to gauge image popularity, often quantified by likes and comments. However, images posted by the same user around the same time may yield drastically different outcomes over time. 
 
To address this variability and enhance the predictive accuracy of image popularity, the concept of intrinsic image popularity ($I^2PA$) was introduced\cite{ding2019intrinsic}. $I^2PA$ focuses on isolating the contribution of visual content to image popularity. The ultimate goal is to understand how the image itself, through its composition, colors, objects, and style, might resonate with viewers and lead to engagement, such as likes, shares, and comments, regardless of external factors like captions, hashtags, the user’s influence, the platform it is shared on, or current trends. 
 
To facilitate the assessment of intrinsic image popularity, researchers have established a large-scale dataset of 2.5 million “popularity-discriminable image pairs” (PDIPs)\cite{ding2019intrinsic}. These PDIPs consist of pairs of images (Image A and Image B) from the same user, posted more than one month and within ten days, captioned with a maximum of six words and have the same hashtags/@ signs. Image A inherently possesses a higher likelihood of popularity compared to Image B by a discernible margin. These constraints ensure that images in each pair have similar textual and social contexts, in order to exclude the external factors other than their visual content, thereby allowing for effective comparison. 

A Deep Neural Network (DNN)-based computational model for $I^2PA$ was then trained using pairwise learning-to-rank approaches. ResNet-50\cite{he2016deep} architecture was adopted as default DNN structure, with the last layer replaced by a fully connected layer yielding a single output representing the predicted popularity score.

\subsubsection{Processing Dataset Images}

In this study, we employed the model for $I^2PA$ in our research to fulfill our objective of predicting the general likelihood of whether an audience would find an image attractive, irrespective of when, where, and who posts it. The obtained scores corresponded to our actual perception. 

\noindent\textbf{\emph{Preprocessing images.}}
Prior to image evaluation, we preprocessed the dataset to align with this study’s purpose. When being arranged in nine-grid layout, each image is appeared as a thumbnail in the shape of a square. To ensure dataset representativeness in this case, we cropped each image to its maximum square dimensions. Vertically elongated images were cropped from the top-left corner, while horizontally elongated ones were cropped from the center. Additionally, each square was resized to a 300px × 300px thumbnail to minimize evaluation errors. 

\noindent\textbf{\emph{Data evaluation.}}
The image combination in every original post is comprised of nine thumbnails. They were grouped together before evaluation. Utilizing the Intrinsic Image Popularity Assessment Model, each thumbnail was assigned a score representing its visual appeal. Scores were ranging from -5 to 5, with higher values indicating greater potential to go viral. 

\noindent\textbf{\emph{Rearranging thumbnails.}}
Within each combination group, thumbnails were ranked based on their assigned scores. Subsequently, two distinct layouts were generated: one following sequential order and the other adhering to the center prioritization principle. In the sequential order layout, the thumbnail with the highest score occupied the top-left corner, proceeding in reading order, with the lowest-scored thumbnail placed in the bottom-right corner. By contrast, in the center prioritization layout, the highest-scored thumbnail occupied the central position, with the subsequent four thumbnails positioned in the four corners, and the remaining four arranged in four lines. Each group resulted in two unique combinations, each formatted into a 900px × 900px image. 

\subsection{Aesthetics-quality Model (NIMA Model)}

\subsubsection{Introduction to NIMA Model}

The Neural Image Assessment (NIMA) model represents a significant advancement in the automatic evaluation of image aesthetics and quality using deep learning techniques, specifically convolutional neural networks (CNNs). Unlike traditional methods that rely on average opinion scores, NIMA aims to emulate human subjective judgments, providing an automated and objective assessment tool for enhancing image content on platforms such as social media.

The core of the NIMA model lies in its utilization of CNNs to extract intricate features from images. These features are then mapped to a continuous score space, reflecting the aesthetic appeal and quality of the images. In contrast to conventional quality assessment methods, NIMA's scoring system delves into the nuanced aspects of image aesthetics, contributing to a more refined evaluation process.

NIMA's evaluation of image aesthetics and quality encompasses several dimensions:
\begin{enumerate}
    \item Technical Quality: This dimension includes parameters such as exposure, color balance, and sharpness, reflecting the fundamental quality and clarity of images. These parameters serve as crucial indicators in assessing overall image quality.
    \item Aesthetic Quality: Encompassing composition, emotional expression, narrative elements, and more, this dimension influences the visual appeal and artistic aspects of images, making it essential for evaluating aesthetic quality.
\end{enumerate}

In the evaluation process, NIMA integrates information from these dimensions and conducts comprehensive and detailed analyses of images using deep learning algorithms, resulting in an objective assessment of image aesthetics and quality. This holistic approach enhances NIMA's applicability and practicality in the field of image evaluation.
The design principles underlying the NIMA model involve:
\begin{itemize}
    \item CNN Structure: NIMA adopts established CNN architectures such as VGG16 or Inception-v2, comprising multiple convolutional layers, pooling layers, and fully connected layers for effective feature extraction.
    \item Feature Extraction: Through successive convolutional and pooling operations, CNNs extract rich features from input images, capturing both local and global information crucial for assessing image aesthetics and technical quality.
    \item Score Distribution Prediction: Unlike traditional models that predict single average scores, NIMA predicts the probability distribution of scores, providing a more nuanced understanding of human judgments and yielding comprehensive evaluation results.
    \item Loss Function: Training NIMA involves defining an appropriate loss function, such as Kullback-Leibler (KL) divergence, to measure the difference between predicted score distributions and human rating distributions, ensuring the model learns accurate representations.
\end{itemize}

In conclusion, NIMA's integration of CNNs for feature extraction and score distribution prediction enables comprehensive evaluation of image aesthetics and quality. Its design principles and methodologies make it a valuable tool for automated and objective image assessment, with applications across various domains including social media content enhancement and image analysis.

\subsubsection{Scoring Process for Dataset Images}

In our study, we utilized the NIMA model to score images from various datasets. These datasets encompass a wide range of image types, themes, and visual characteristics, providing a comprehensive scope for evaluating the performance of the NIMA model across different contexts. Standardized scoring criteria were employed, where higher scores indicate higher aesthetic appeal and image quality.

The scoring process involved the following steps:
\begin{enumerate}
    \item Data Preprocessing: Prior to scoring, we preprocessed the images in the datasets to ensure they met the input requirements of the NIMA model. Preprocessing steps include resizing images, color correction, and noise removal.
    \item Scoring Calculation: Using the trained NIMA model, we calculated scores for each image. The NIMA model outputs a continuous score distribution, reflecting the image's potential performance in terms of aesthetic appeal and quality.
    \item Result Interpretation: Based on the NIMA model's output, we interpreted and recorded the scores for each image. These scores reflect the level of perceived attractiveness and quality of the images, providing foundational data for subsequent image arrangement strategies.
\end{enumerate}

The evaluation results from the NIMA model serve as fundamental indicators for subsequent image arrangement strategies. Guided by these scoring results, we directed the arrangement of images within a nine-grid layout.

\subsection{Rearranging Images}
After obtaining scores from both the Aesthetics-Quality Model and the Intrinsic Image Popularity Assessment Model, we proceeded to rearrange the images in the nine-grid layout. Each model's scores were used to determine the order of arrangement, resulting in four distinct schemes: Aesthetics-Quality Model-based sequential arrangement, Aesthetics-Quality Model-based center prioritization arrangement, Intrinsic Image Popularity Assessment Model-based sequential arrangement, and Intrinsic Image Popularity Assessment Model-based center prioritization arrangement.

In the Aesthetics-Quality Model-based sequential arrangement, thumbnails were ordered in a reading order starting from the top-left corner, with the highest-scored thumbnail positioned first and the lowest-scored thumbnail placed in the bottom-right corner. This sequence aimed to maintain a logical flow for viewers' visual perception.

Conversely, in the Aesthetics-Quality Model-based center prioritization arrangement, the highest-scored thumbnail took the central position, highlighting its significance. The following four thumbnails were placed in the four corners surrounding the central image, while the remaining four thumbnails were arranged in four lines around the central block. This layout intended to emphasize the highest-rated image while ensuring a balanced distribution of other images around it.

Similarly, the Intrinsic Image Popularity Assessment Model-based sequential arrangement and center prioritization arrangement followed the same principles but were based on the scores from the Intrinsic Image Popularity Assessment Model.

Each group resulted in two unique combinations, each formatted into a 900px × 900px image. These meticulously designed arrangements serve to optimize the visual impact and aesthetic appeal of image sets within the nine-grid framework. By providing data for subsequent user annotation, these arrangements lay the groundwork for further research into which model-based arrangement best aligns with user preferences.

\begin{figure}[!t]
    \centering
    \begin{subfigure}[b]{0.4\textwidth}
        \includegraphics[width=\textwidth]{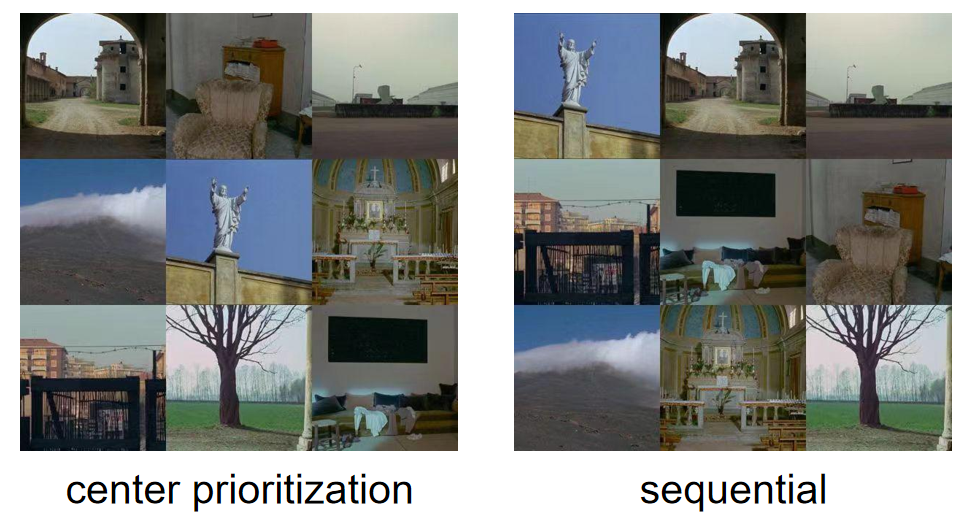}
        \caption{An example of rearrangement based on NIMA model.}
    \end{subfigure}
    \hfill
    \begin{subfigure}[b]{0.4\textwidth}
        \includegraphics[width=\textwidth]{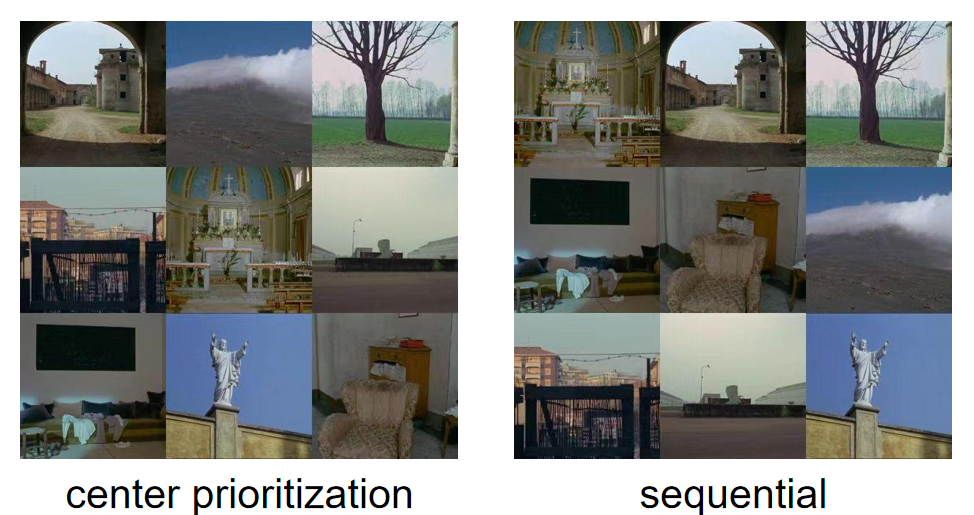}
        \caption{An example of rearrangement based on \textit{I\textsuperscript{2}PA} model.}
    \end{subfigure}
    \caption{Examples of image rearrangements.}
    \label{fig:rearrangements}
\end{figure}

\section{Study Design}

\subsection{Dataset}
Currently, there is not much research on nine-grid layout, so there are currently no publicly available nine-grid datasets. Therefore, we need to build the dataset ourselves, and the nine-grid layout in this dataset must be recently released. We crawled a total of 1,901 nine-gram charts published by 21 users on the sina weibo client as a dataset. These users come from a variety of fields, including film, design aesthetics, humanities, photography, stage design, and more. All of these users have more than 2,000 followers, including 17 users with more than 10,000 followers, 16 users with more than 100,000 followers, and 7 users with more than 1,000,000 followers. We chose users with more followers because users with more followers are likely to be more concerned about the quality of the nine-grid layouts they post, thus reducing the likelihood of getting low quality nine-grid layouts. We selected 250 of these nine-grid layouts that are representative of each domain . We arranged these nine-grid layouts in order of highest to lowest center layout according to aesthetic rating, highest to lowest order layout according to aesthetic rating, highest to lowest center layout according to content attractiveness and image quality, and highest to lowest order layout according to content attractiveness and image quality, so that the nine images in each original nine-grid layout form four layouts.

\subsection{Data Collection}
In order to test user preferences for the four nine-grid layouts, we used a publicly posted questionnaire to collect users' preferences for the nine-grid images. The users who participated in the questionnaire were drawn from the entire community. We used a four-alternative approach, where users were asked to continuously choose their favorite arrangement from four different arrangements of the same nine-box grid.In order to reduce the interference of the position of the options, we designed different layout schemes for different questionnaires. Also to prevent user fatigue, we created 5 questionnaires with 50 questions in each questionnaire, and users need to answer all 50 questions before submitting the questionnaire.
We received a total of 45 valid questionnaires, and the vote counts for the four arrangement schemes are presented in the table.

\begin{figure}
        \centering
        \includegraphics[width=0.4\textwidth]{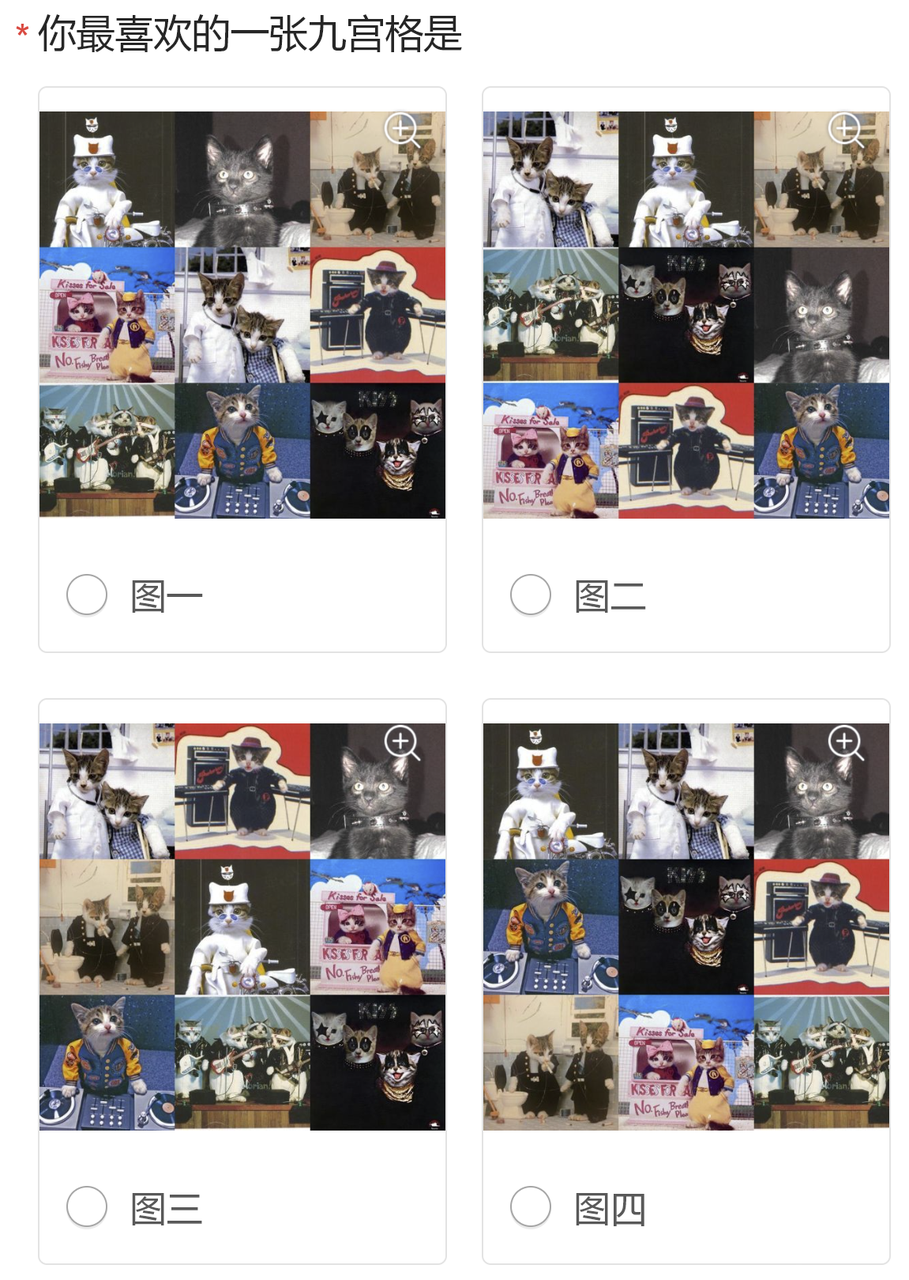}
        \caption{Subjects were asked to choose their favorite of four permutations given}
        \label{fig:enter-label}
\end{figure}

Overall, the arrangement scheme based on aesthetic quality outperformed the one based on visual content of the images, while the principle of center prioritization received higher evaluations compared to the sequential arrangement. 
Specifically, the combination of aesthetic ratings for individual images and the center prioritization sequence yielded the best results, whereas the scheme of sequentially arranging the images based on visual content ratings was noticeably inferior to the other schemes. Furthermore, both the sequential arrangement based on aesthetic ratings and the center prioritization arrangement based on visual content received favorable evaluations.

\begin{table}
    \centering
    \begin{tabular}{c|c|c}
    {}&Aesthetics-Quality & Visual content\\
    center prioritization&628&585\\
    sequential & 599 &438\\
    \end{tabular}
    \caption{Results of the vote on the best option}
    \label{tab:my_label}
\end{table}

\subsection{Result}
From the previous formative research, we have learned that although the first and center positions in a grid display are noticeably more likely to be prioritized over other positions, the center position also receives significantly higher attention than the first position. The statistical results from user annotations further support the evidence that a center prioritization arrangement scheme aligns with user preferences. This conclusion contributes to the understanding of human visual attention mechanisms and provides information for the development of theories and models related to visual attention and perception in psychology.

The process of human image perception is complex, as various factors such as color, brightness, clarity, subject, and background can influence people's evaluations. Furthermore, different individuals may have different evaluations of the same image, which implies that the ratings provided by models may not be entirely accurate, leading to potential errors.In our experiment, we utilized the aesthetic quality scores provided by the NIMA model and the content ratings provided by the I2PA mode,both models demonstrated high accuracy and performance in their datasets. Therefore, we can argue that aesthetic appeal and quality of an image are more closely aligned with the way people evaluate images compared to all the visual information present in the image.

\section{Discussion}
\noindent\textbf{\emph{Scope of our study.}} The nine-grid layout is prevalent in social media, user interfaces, game design, and information display. The arrangement of images in this layout significantly impacts user experience, efficiency, and aesthetic appeal. Previous research has focused on image popularity, quality, and the impact of positioning on engagement, but the effect of image order on user preference in a nine-grid layout is underexplored. Our study investigates the impact of different arrangement methods on the visual appeal and narrative expression of nine-grid images, providing guidance on optimal arrangement strategies. We consider the serial-position effect, emphasizing recall of the first and last items, and visual attention distribution, suggesting central focus, to determine which view is more attractive in the nine-grid layout.

\noindent\textbf{\emph{User's cognition of nine-grid images.}} Through questionnaires and interviews, we found that users post multiple pictures to document and share their lives. They select images based on aesthetic appeal, thematic relevance, and technical quality, adjusting the order for visual effect. Both the center and upper-left positions are important in a nine-grid layout. Users are interested in arranging images to create captivating posts, suggesting the need for ranking tools, templates, and arrangement suggestions to enhance satisfaction and engagement.

\noindent\textbf{\emph{The Utilization of Models.}} We used the Aesthetics-Quality Model and the Intrinsic Image Popularity Assessment Model to score images and rearranged them in a nine-grid layout using four schemes. The center prioritization scheme places the highest-scored image in the center, while the sequential arrangement follows a logical flow from top-left to bottom-right. This processed dataset provides data for user annotation studies, supporting further research into user-preferred layouts.

\noindent\textbf{\emph{User’s preference of nine-grid layout.}} Statistical data from user labeling experiments show that the center position is crucial, and the center-first layout aligns with user preferences. Models incorporating aesthetic quality more accurately reflect social platform evaluations. Enhancing model performance can further improve the effectiveness of arrangement schemes.

\subsection{Future Work}
\noindent\textbf{\emph{Optimizing the nine-grid layout.}} Future research should include long-term user behavior analysis to understand the sustained attractiveness and effectiveness of different arrangements in the nine-grid layout. Exploring user preferences, customization, and cross-cultural studies will help tailor designs to individual needs and identify universal or culture-specific attractiveness factors. Investigating dynamic, interactive, and multi-model elements will contribute to engaging and visually appealing user experiences, enhancing the nine-grid layout's attractiveness and providing practical design optimization guidance.

\noindent\textbf{\emph{Generative tool for nine-grid layout on social media.}} We developed a Generative tool that automates content arrangement, offering optimal placement suggestions based on visual attractiveness, thematic coherence, and user engagement. This tool streamlines workflows for social media managers and content creators, enhancing the aesthetic appeal and user experience of social media profiles. With customization options and creative variations, the Generative tool enables users to design captivating nine-grid layouts effortlessly.

\section{Conclusion}
In this study, we explored ways to make multi-image posts on social media more appealing and engaging. By using computational models to assess image aesthetics and content attractiveness, we aimed to optimize the arrangement of images within the nine-grid layout.
Our findings suggest that arranging images based on aesthetic quality and prioritizing central positions leads to more favorable user evaluations. This research provides valuable insights for content creators looking to enhance the visual presentation of their posts on social media platforms, ultimately improving user engagement and overall content experience.

\bibliographystyle{ieeetr}
\bibliography{references}

\end{document}